\documentstyle[12pt, fleqn]{article}
\begin{document}
\baselineskip 9mm
\

\begin{flushright}
AZPH-TH/93-5
\end{flushright}
\begin{center}

{\bf Stochastic Fluctuations and Structure Formation in the Universe}

\bigskip
\vspace{20mm}

Arjun Berera$^1$ and Li-Zhi Fang$^{1,2}$

\bigskip

$^1$Department of Physics\\
University of Arizona\\
Tucson, AZ 85721\\

\bigskip

$^2$Steward Observatory\\
University of Arizona\\
Tucson, AZ 85721\\

\vspace{20mm}
\end{center}

\newpage
\begin{center}
{\bf Abstract}
\end {center}

It is shown that the evolution of the density perturbations
during certain
eras of substantial entropy generation in the universe can be
described in the scheme of the
KPZ equation. Therefore, the influence on cosmological
structure formation by stochastic forces arising from
various dissipations can be
studied through the universal characteristics of surface growth in
$d=3+1$ dimensions.
We identify eras of strong stochastic fluctuations and
describe dynamically how these other dissipative sources of noise, besides
initial (inflationary) quantum fluctuations, generate seeds of
density perturbation with
power law spectrum, including the Harrison-Zeldovich
spectrum.
\vspace{20mm}

PACS numbers: 98.80.Bp, 98.65.Dx, 05.70.Ln

\newpage

\bigskip

In the last few years, significant progress has been made in
understanding the dynamics of growing a rough or structural
surface from an initially flat surface by random fluctuation
\cite{vicsek}.
Many structure formations in physics have been understood by
studying the scaling properties of their growth patterns.
Central to these studies has been the characterization
of universal properties associated with systems of diverse
physical attributes.
To guide in these investigations, a major breakthrough has been the
development of systematic analytic treatments
inspired by scaling and renormalization group theory \cite{kardar}.
This treatment, aimed at studying the spatial and temporal behavior of
structural growth, has revealed that the universal
scaling properties come from the non-linear and stochastic terms in
the dynamical equation.

In principle, the structure formation in the universe can also be
classified as the phenomena of structural 'surface' growth. Big Bang
cosmology essentially tries to explain how an initially homogeneous
mass distribution evolved into its present inhomogeneous
state.  In the language of the spacetime metric, it explains how an
initially
flat or smooth 3-dimension surface described by the
Robertson-Walker metric evolved into a wrinkled one.
The analogy to surface formation takes root by associating
the initial mass distribution with a
flat three-dimensional surface and
its subsequent structure formation as that of surface roughening.
This analogy gains interest by noting that
cosmological structure also shows scaling in, for example,
two point correlation
functions of galaxies, clusters of galaxies and quasars, all
of which behave as
$r^{-\gamma}$ with $\gamma \sim 1.8 $ up to present
day scales of about 300 Mpc.

In
the standard inflationary model, it is assumed that the seeds of
the density perturbation are produced by the quantum noise of scalar fields
during the inflation era \cite{kolb}. In this model scaling
structure is therefore explained as due to white
noise seeds from this quantum fluctuation. However, besides quantum
fluctuations, there are also time periods when stochastic fluctuations
are large and which can, as we will see, lead to scaling
seeds.
Although this connection of stochastic
fluctuations to scaling makes it of special interest, it
should be recalled that quite generally dissipations must be
accompanied by fluctuations or stochastic forces.
In cosmology much work has concentrated on the effects
of dissipation, for instance during the reheating period.
This dissipation by the damping of scalar fields must also imply
fluctuations of it, which preempts the investigation
in this paper.

Analytic studies have shown that scaling behavior is common
to systems that obey nonlinear dynamical equations with also a stochastic
driving term \cite{vicsek}.
The structure formation in the universe, in particular at
subhorizon scales, is just such a system.
Therefore, it is worthwhile to study the models of cosmological
structure formation from the point view of the universal dynamics
that governs structural surface growth.
In this paper we will quantify the analogies drawn above and then
focus on the influence of stochastic fluctuations
on structure formation. We clarify that
it is already known structure
formation was predominately at superhorizon scales during the inflation
era and so must be treated by general relativity \cite{bardeen}.
However, we will show that in specific
periods when dissipation becomes significant,
the influence of fluctuation to structure formation is of subhorizon
scales, and
can be described by a non-relativistic equation.

\bigskip
\bigskip

{\noindent} 1. The standard model(s) of cosmic structure formation
({\it e.g.} inflation theory) assumes that the initial spectrum
of density perturbation was given by the vacuum quantum fluctuations
and inflationary expansion, and that the subsequent evolution of
clustering was deterministic, i.e. it obeyed a dynamical equation
{\it without a noise term}. This is equivalent to assuming
that either $a$) no noise sources existed after the inflation era
or $b$) the influence of post-inflation noise on structure formation
was negligible.

Obviously assumption $a$ is not true, because dissipation (or
processes of approaching locally thermal equilibrium) was
essential in the eras of cosmic entropy generation, and generally such
dissipative processes would lead to a stochastic force ${\bf F}$
(fluctuation-dissipation theorem). In the standard model, these
dissipative eras at least included reheating of inflation, baryongenesis,
non-thermal equilibrium decoupling of particles, and post-inflation
phase transitions. Moreover, the actions given by turbulence-like
perturbations and explosions were also essentially stochastic.
One can expect that in such eras cosmic matter was influenced
significantly by the stochastic fluctuation force ${\bf F}$.

Turning to assumption $b$, it is correct if the non-linear terms in
the dynamical equation can be neglected. Without the non-linear term,
the noise force ${\bf F}$ will not change the scenario of clustering
as given by the linear approximation, but only contributes to a
statistical error in the result. However, the influence of noise will
no longer be trivial, as we shall show, if non-linear corrections
to the dynamical equation are considered.

As an example, let us consider the era of reheating after inflation,
during which there was out-of-equilibrium decay of massive,
nonrelativistic particles.
This process can be described as "friction"-like coupling in the
dynamical equation \cite{vicsek}. Since by this time period
causality forbids any new formation of fluctuations at
super-horizon scales, the only
fluctuations raised by the entropy generation
of reheating are of sub-horizon scales.
Moreover, during the period of coherent oscillation,
as the scalar field damps, the universe becomes
matter dominated by these nonrelativistic particles.
Therefore, in this time interval,
the influence of stochastic
fluctuations on structure formation
can be described by the non-relativistic hydrodynamical equation
of structure growth in an expanding universe. In linear approximation, the
momentum equation is given by \cite{kolb}
\begin{equation}
{\frac {\partial {\bf v}}{\partial t}}+
{\frac {\dot{R}}{R}}{\bf v}+
{\frac {\dot{R}}{R}}({\bf r} \cdot \nabla){\bf v}
+ {\frac {v_s^2}{\rho_0}}\nabla \rho +\nabla \phi =0
\ ,
\end{equation}
where the density $\rho$, peculiar
velocity ${\bf v}$ and gravitational potential
${\bf \phi}$ are the perturbations
to the basic-state (smooth) solutions $\rho_0$,
${\bf v_0}$, $\phi_0$. $R(t)$ is the cosmic scale factor and $v_s$
is the speed of sound.
A straightforward examination of the linearized equations
shows that only the vorticity free modes can be amplified by
gravitational instability in an expanding universe.
Therefore, we will only consider those solutions satisfying
the constraint
$\nabla \times {\bf v}=0$.
In this case, one can define a velocity potential $\psi$ by
\begin{equation}
 {\bf v} = - \nabla \psi.
\end{equation}

On the other hand, it is well known that in linear approximation
the velocity ${\bf v(r},t)$ is proportional to the gravitational force
produced by the surrounding density perturbation.
Thus we have the local relation
\begin{equation}
\rho = - f\nabla \cdot {\bf v},
\end{equation}
where $f=4\pi\rho_0/H_0$ in a flat ($k=0$) universe. From eqs.(2)
 and (3),
one has $\rho = f\nabla^{2} \psi$. Therefore, $\psi$ is
proportional to the gravitational potential by the relation
$\phi = 4\pi Gf\psi$, so that
${\bf v}=  -(4\pi G f)^{-1} \nabla \phi$.
Substituting eqs.(2) and (3) into eq.(1), one has
\begin{equation}
{\frac {\partial {\bf v}}{\partial t}} +
{\frac {\dot{R}}{R}}{\bf v} +
{\frac {\dot{R}}{R}}({\bf r} \cdot \nabla){\bf v}
= {\frac {v_s^2}{\rho_0}} f \nabla^2{\bf v}
+ 4\pi Gf {\bf v}
\end{equation}
This equation is similar to the Langevin equation but without a
stochastic force. The first term on the right-hand side of eq.(4)
describes
relaxation of the structure by diffusion. The second term formally
corresponds to the viscosity term in the Langevin equation, but
here the sign is negative, because self-gravitation leads to acceleration,
not deceleration of the clustering matter.

As discussed above, during the eras of dissipation in the universe,
the dynamical equation (4) should include a
stochastic force or noise term, ${\bf F}$, on the right side.
Then eq.(4) finally has the form of a Langevin-like equation.
The stochastic force acting on the vorticity-free perturbation
should be
\begin{equation}
{\bf F}=\nabla \eta({\bf x}, t)
\ , \end{equation}
where the noise $\eta({\bf x}, t)$ satisfies
$<\eta({\bf x}, t)>=0$. If the noise is Gaussian, we have
$<\eta({\bf x}, t)\eta({\bf x}, t)>=
2D\delta^{3}({\bf x}-{\bf x'})\delta(t-t')$, where
$D$ is the mean square variance of the noise. More generally, the
spatial-temporal Fourier transform of $\eta({\bf x},t)$ satisfies
\begin{equation}
<\eta({\bf k},\omega)\eta({\bf k'},\omega')>
=2Dk^{-2\chi}\omega^{-2\theta}\delta({\bf k+k'})\delta(\bf \omega+\omega')
\ , \end{equation}
where for the case of Gaussian noise $\chi=\theta=0$.

At linear approximation as we are considering in eq.(4),
the solution will essentially not
be affected by the stochastic force ${\bf F}$, because
the noise term  can be eliminated from the dynamical equation
 upon averaging, regardless of the value of $D$.
As such, the noise term simply leads to an increase of statistical
 variance in the linear results.

However, adding non-linear corrections to eq.(4) will substantially
change this scenario.
The lowest order non-linear correction of eq.(4) is given
by the Euler
term $({\bf v}\cdot\nabla){\bf v}$.
Including this and the noise term,
eq.(4) has the modified form,
\begin{equation}
{\frac {\partial {\bf v}}{\partial t}}
+{\frac {\dot{R}}{R}}{\bf v}
+{\frac {\dot{R}}{R}}({\bf r} \cdot \nabla){\bf v}
+ ({\bf v}\cdot\nabla){\bf v}
= {\frac {v_s^2}{\rho_0}} f \nabla^2{\bf v} +
4\pi Gf {\bf v}+ {\bf F}
\ .
\end{equation}
Strictly speaking, we should also include the dissipative term
corresponding to ${\bf F}$ in eq.(7). However, this linear term will
not affect the main results discussed below.

Eq. (7) governs the evolution of matter perturbation
during the beginning period of reheating when the scalar field
is undergoing coherent oscillations. Obviously, eq.(7) is not limited
to this period of reheating in the early universe,
but generally describes the
behavior of structure formation in any era when,
1) dissipation is significant, and 2) the universe is dominated by
non-relativistic particles. Other possible examples, besides the reheating
case, are
late-time phase transitions \cite{witten} and
collision or merging of galaxies.

If the interaction causing the stochastic force is weaker
than self-gravitation, and/or its time scale is comparable to or even
longer than Hubble
expansion, the noise term will be less important. One can call this the
case of weak noise. For instance, the stochastic force related to the
bulk viscosity at last scattering surface \cite{weinberg}
is negligible, because the
entropy per baryon was very large at the era of last scattering.

However, for the eras in which the main or a comparable part of
cosmic entropy was generated, as
in the part of the reheating period discussed
above, dissipation would have been crucial,
even dominant in the evolution of the universe \cite{fang1}.
Therefore the relevant stochastic force ${\bf F}$ would have been stronger
than self-gravitation and its time scale less than that of
Hubble expansion. In such periods, one can neglect the
cosmic expansion ($R(t)$) and self-gravitation
($4\pi Gf {\bf v}$) terms. Eq.(7) then becomes the Burger's
equation \cite{burgers} with stochastic force
\begin{equation}
{\frac {\partial {\bf v}}{\partial t}}
+ ({\bf v}\cdot\nabla){\bf v}
= {\frac {v_s^2}{\rho_0}} f \nabla^2{\bf v}+{\bf F}
\ .
\end{equation}
It has already been recognized that the non-linear evolution
of cosmic density inhomogeneities can be approximately treated by
the Burger's equation \cite{gurbatov}.  This this work concentrated on the
the formation of pancakes and filaments, and did not
encorporate stochastic
forces, which are central to our considerations.

Using eq.(3) and eq.(8) one finds the equation for
$\phi$ to be
\begin{equation}
{\frac {\partial \phi}{\partial t}}=
f'\nabla^2 \phi + {\frac {1}{2}}({\nabla \phi})^2 + \eta({\bf x}, t)
\ , \end{equation}
where $f'= (v_s^2/\rho_0) f$.
Both the above equations
are variants of the
so called KPZ equation \cite{kardar},
which has been widely studied as a dynamical equation for describing
universal behavior of fractal surface growth under
stochastic force.
{}From our considerations
we see that the influence of stochastic forces on cosmological
clustering also belongs, under
certain approximations, to the dynamics given by the KPZ equation.
If one makes an analogy with the theory of surface growth,
one finds that the gravitational
potential $\phi$ of cosmic matter corresponds to the height of
the surface.
This means that the evolution of the gravitational
potential undergoing stochastic fluctuation is analogous to the
problem of d=3+1 surface growth, i.e. a 'surface' growing on a
3-dimension substratum.

\bigskip
\bigskip

{\noindent} 2. Eqs.(8) and (9) show that
noises in strong dissipation eras would input the corresponding
scaling seeds into the mass distribution, which
subsequently would then be amplified by  gravitational instability in
the expanding universe. As such, stochastic
forces from strong dissipation eras would most likely leave
some signatures in today's clustering.
In order to illustrate the influence of the noise on the clustering,
let us turn to the correlation functions.
The seeds generated by noise normally are scaling with correlation
functions going as,
\begin{equation}
<\phi({\bf x},t)\phi({\bf x'},t)> \sim \mid {\bf x} -{\bf x'}\mid^{\alpha}
\ ,
\end{equation}
so that the two-point correlation function of density is then
\begin{equation}
\xi(r) = <\rho({\bf x},t)\rho({\bf x'},t)>\sim r^{-\gamma}
\ , \end{equation}
where $\gamma=4-\alpha$.

The index $\alpha$ depends on the spectrum of the noise in eq.(6).
Unfortunately, for $d=3+1$, few firm relationship between ($\chi$, $\theta$)
and $\alpha$ are available.
However, one can find the possible range of $\alpha$ from the
following universal relation \cite{vicsek}
\begin{equation}
\alpha=4\beta/(\beta + 1)
\ , \end{equation}
where $\beta$ is the index for the time behavior of the correlation function
of $\phi({\bf x}, t)$  at a given ${\bf x-x'}$,
\begin{equation}
 <\phi({\bf x},t)\phi({\bf x'},t)> \sim t^{2\beta}
\ .
\end{equation}
If we examine the case of perturbations which grow faster than
the gravitational instability, it would require
$\beta$ to
be larger than $1/2$ in the radiation era or $2/3$ in the matter era.
This would mean that we have, respectively
\begin{equation}
0<\gamma< 2.66 \ or \ 2.4
\ , \end{equation}
thus indicating how the effects of
noise, or more generally the dynamics embodied in
eqs. (7), contribute nontrivially to structure formation.
It also means the basic assumptions $a$ and $b$,
implicit to the standard (inflation) model, are not necessarily true.

One can further quantify these results by using the following
general relation, obtained by perturbative methods in \cite{medina},
\begin{equation}
\alpha=(4\chi-2d+8\theta+6)/(2\theta+3)
\ .\end{equation}
which is valid for  $0<\chi<2$ and $0<\theta<0.25$.
Within the limits that one accepts this perturbative result
to give a semiquantitative guide, one can obtain relations
between $\alpha$ and the spectrum of spatial ($\chi$) and
temporal ($\theta$) noise.  For example, one can obtain the
solution in the near proximity of the observed two-point
density correlation function $\alpha \approx 2.3$
($\gamma \approx 1.8$) for $\chi \approx 2$ and
$\theta \approx 1/6$.  This is a suggestive example
since it has the following interpretation.
{}From eqs.(3) and (5) observe that $\nabla^2 \eta$ is proportional
to the stochastic force term acting on the
density fluctuation $\delta \rho$, so that this solution
corresponds to a white noise fluctuation on the density
but with a temporal correlation.  The latter should not
be surprising since dissipative eras are of finite
temporal extent and so would reflect on the temporal
noise correlation.
The above  demonstrates one way that a strong white noise in
the early universe would be able
to generate an initial perturbation
which along with possible further modifications
by gravitational instability could leave signatures in present
observation.
Also contained within the solutions of (15) is $\alpha=0$,
which corresponds to the Harrison-Zeldovich spectrum. Finally,
 eq.(15) tentatively shows that $\alpha$ increases with $\chi$.
As a reasonable extrapolation, assuming that such a trend holds for
$\chi >> 2$, it is
suggestive that such strong, turbulence-like noise may not be consistent
with observation. Of course, we should keep
in mind that eq.(15) is a perturbative result, so that this
deduction is not rigorous and also may not be
unique since nonperturbative results
could exist.

\bigskip
\bigskip

Standard model cosmology has eras, such as during reheating,
when stochastic fluctuations given by cosmic phase transitions and
other non-thermal equilibrium processes are significant.
During such times, their effect may play a
non-negligible role in structure formation.
The last conclusion was also reached by Luo and Schramm \cite{luo}
based on observational data indicating
scale-free distributions of galaxy clusters.  They concluded
the need for encorporating a fractal structure generation mechanism
into standard big bang cosmology.  Although, as pointed out
by Peebles \cite{peebles}, a pure fractal contradicts the observed
large scale angular correlation function, their
considerations were restricted to subhorizon scales
of order 300 Mpc in the present day universe.
They explained the mechanism based on an aggregate growth
process and concluded that fractal growth originates
from two-dimensional sheetlike
objects.  However, as we have shown, the dynamical mechanism
is rather more general and is governed by the KPZ-equation,
which is not necessarily restrive to two dimensional
growth phenomenon but rather to "surface" growth in
higher dimensions also.  Within the framework of the
formalism presented here,  we learn that the
universal characteristics of rough surface growth, such as
the relationship between the noise spectrum and the index of
the two-point correlation function, can be used for guidance in
developing models of structure formation in the universe. This provides
a two-step approach for checking models in particle cosmology: (a) testing the
standard model by calculating the contributions of stochastic
fluctuations related to various dissipations in the universe; (b)
testing particles physics models which may give rise to stochastic forces
in the early universe by assuming the correctness of the standard model.

\bigskip
\bigskip

We thank Professor Y. C. Zhang for helpful discussions.
Financial support was provided to Arjun Berera by the
U.S. Department of Energy, Division of High Energy and Nuclear Physics.

\end{document}